\documentclass[10pt,a4paper]{article}

\usepackage{amsmath,amssymb}
\usepackage{graphics,graphicx}
\usepackage{epsfig}
\makeatletter
\@addtoreset{equation}{section}
\makeatother

\begin{document}
\begin{flushright}
QMUL-PH-05-14\
\end{flushright}
\begin{center}
\Large\textbf{Instantons, Assisted Inflation and Heterotic M-Theory.}\\
\vspace{1.2cm}
\normalsize
\textbf{John Ward}\footnote{j.ward@qmul.ac.uk}\\
\vspace{0.8cm}
\emph{Department of Physics \\ Queen Mary, University of London\\ Mile End Road, London\\E1 4NS. U.K.}
\end{center}

\begin{abstract}
We first investigate the attractor solution associated with Assisted Inflation in Heterotic M-Theory to see if it is stable. By perturbing
the solutions we find a solitary stable fixed point with the general path through phase space dependent upon the Calabi-Yau data and the number of five-branes.
We then go on to examine the effect of including non-perturbative corrections to the inflaton potential arising from boundary-brane instantons as well as
higher order brane interaction terms.
The result is that Assisted Inflation is only possible if we allow fine tuning of the superpotential, and that generic non-perturbative superpotentials
will prevent this kind of inflation from occurring.
\end{abstract}

\section{Introduction.}
The past few years have seen renewed interest in the embedding of realistic cosmology into string theory. Using the ever increasing amount
experimental data, we have been able to impose serious constraints upon the types of string motivated models that could exist.
Most of this work had been concerned with compactifications of type IIB string theory using Calabi-Yau orientifolds.
In particular the seminal $\mathbb{K}LT$ paper \cite{kklt} showed it was possible to obtain de-Sitter vacua by stabilising the complex structure and the dilaton using fluxes wrapping various cycles of the Calabi-Yau. The volume modulus can then be stabilised using a constant superpotential.
However once stabilisation of the moduli has been achieved, it is still non-trivial to find inflationary behaviour from $F$-terms preserving $\mathcal{N}=1$ supersymmetry due to the $\eta$ problem, where the slow roll parameter $\eta$ is always of order unity implying too large a mass for the inflaton.
This has been the stumbling block in recent years with regard to string theoretic approaches. Furthermore the flux stabilisation techniques only seem to yield
Anti de-Sitter vacua, which are not physical. The uplift to de-Sitter space can be accomplished by turning on matter fields on the $D3$ or $D7$-branes \cite{fluxes}
wrapped on various cycles \cite{comment}.
Recent work has also demonstrated how to uplift to de-Sitter vacua in M-theory \cite{desitter, moduli_stabilisation}, and so it is natural to enquire as to whether the same problems arise here.

The apparent difficulty in realising inflation in M-theory is due to the fact that the superpotentials are generally exponentials arising from various non-perturbative
effects \cite{buchbinder, copeland, decarlos, witten, lukas} and therefore lead to steep inflaton potentials.
This suggests that embedding inflation in M-theory will be difficult as the moduli fields can not easily be stabilised.
In the standard inflationary scenario it is necessary to have potentials with at least one flat direction in order for the inflaton field to slowly roll
toward the vacuum \cite{becker}, and thus generate the required number of e-foldings in line with experimental constraints. The various other moduli must have strongly curved
potentials in order for them to stabilise, and they must be stabilised before inflation can occur. The exponential potentials in M-theory can be advantageous for inflation
as the potential in the inflaton direction will generally be the steepest, thus there is no need to stabilise the various other moduli as they will evolve much slower than
the inflaton. The only issue in this context is whether it is possible to ensure that the inflaton will roll sufficiently slowly to allow the universe to inflate,
whilst simultaneously evolving fast enough so that we can neglect the dynamics of the other moduli. One possible way to satisfy this is through Assisted Inflation \cite{assisted_inflation}, 
whereby we consider several scalar fields which can individually have steep potentials. The key point is that
it is the combined effect of the fields that appears in the friction term in the equation of motion, thus more fields implies more friction.
This is precisely what allows us to satisfy the e-folding constraints and therefore a model compatible with current data. Furthermore because we are considering compactification on
a Calabi-Yau space, the equations of motion for the scalar fields have direct geometrical interpretation \cite{isometries} allowing us to use geometrical methods to
extrapolate a large number of cosmological solutions..
In String/ M-theory this type of scenario can be realised either by combining scalar fields such as the tachyon \cite{tachyon_inflation}, or by considering a large number of branes - where
we use the inter-brane distance to play the role of the inflaton. 
This was analysed in the context of Heterotic M-theory in a recent paper \cite{becker}, where the dynamics of several five-branes along the orbifold was mapped to Power Law Inflation \cite{powerlaw}. It was shown that the number of 
scalar fields (and therefore branes) required for inflation was of the order $N \sim 89$ in order to satisfy cosmological constraints. Of course it should also be noted
that generally scale factor expansion differing from an exponential tends to tilt the metric fluctuation spectrum away from being scale invariant.

Another benefit of realising inflation in this context is that the standard model can live upon one of the boundaries of the orbifold interval, whereas
in type IIB theory it has been argued that a separate throat is required in the Calabi-Yau where the standard model lives.
This has serious implications for reheating, as we need to ensure the graviton wavefunction links the inflationary throat and standard model throat
strongly enough to channel all the inflationary energy into standard model degrees of freedom. Recent work has illustrated that multi-throat scenarios may
be more useful in this context \cite{shiu}. In any case the inflaton sector does not appear to directly couple to the standard model sector, however in Heterotic M-theory
with the standard model degrees of freedom lying on the orbifold fixed point, we expect reheating to occur through small instanton transitions whilst supersymmetry will be broken and the 
moduli stabilised by gaugino condensation when the five-branes strike the boundary branes. Although the exact mechanism for this is still unclear. From a phenomenological perspective, it appears there are many
interesting things to discover about inflation and five-brane dynamics in this set-up.

Despite the relative success we suspect that the Assisted Inflation scenario \cite{becker} is a special case, 
and that this form of inflation will not generally occur in Heterotic M-Theory.
In this note we will test this statement by considering higher order instanton contributions to the superpotential. For example we know that the initial conditions
for inflation will depend heavily upon the inter-brane instanton contributions and so it is worthwhile to study how they affect the evolution of the scalar field and
consequently the dynamics of the five-branes.
We begin this note with a brief review of the Assisted Inflation scenario in Heterotic M-theory \cite{becker}. We then study the phase space analysis of such a model before
moving on to consider various corrections arising from higher order instanton contributions. We will close by summarising our results and suggesting possible avenues
of future research.

\section{Assisted Inflation from M-Theory.}
In this section we will briefly review the Assisted Inflation model with $N$ five-branes in Heterotic M-Theory as introduced in \cite{becker}.
We begin with eleven dimensional supergravity theory on the orbifolded space $\mathbb{S}^1/\mathbb{Z}_2 \times \mathbb{M}_{10}$,
which is coupled to the two 10-dimensional $\mathbb{E}_8$ SYM theories located at the orbifold fixed points.
If we now compactify this on a Calabi-Yau three fold, the resultant theory becomes $\mathcal{N}=1$ supersymmetric in 5-dimensions. The two orbifold fixed points
now each comprise an $\mathcal{N}=1$ gauge theory in 4-dimensions living on the world volume of three-branes, and they are separated along the fifth dimension.
This is more commonly referred to as the visible and hidden sector of the theory.
What is useful for our description is the fact that we can additionally place $N$ five-branes along the transverse direction to the fixed points which
will survive the compactification down to five-dimensions, and it is the dynamics of these branes along the orbifold direction will generate inflation.
Note that the Assisted Inflation scenario will only be valid if the branes are all oriented in the same direction and are parallel to the boundary branes. In some
sense this is one of the 'BPS' configurations \cite{buchbinder2}.
We will also take these branes to wrap the same holomorphic genus zero two-cycle on the Calabi-Yau \footnote{For simplicity we will 
demand that they wrap this cycle only once.},
leaving us with a $3+1$ dimensional theory. This is useful for constructing cosmological models as we can assume that the standard model lives on one of the boundary branes,
and thus the dynamics of the five-branes along the orbifold direction corresponds to the inflaton sector.

The effective four dimensional theory contains various real moduli fields. Firstly there are the two moduli $S$ and $T$, where $S$ describes
the volume of the Calabi-Yau and $T$ is associated with the length of the orbifold. The relative positions of the $N$ five-branes with respect to one of the orbifold
fixed points are given by $Y_i$. For completeness we write the full definitions of the various moduli below.
\begin{eqnarray}
S &=& \mathcal{V} + \mathcal{V}_{OM} \sum_{i=1}^{N} \left(\frac{x_i^{11}}{L} \right)^2 + i\sigma_s, \\
T &=& \mathcal{V}_{OM} + i \sigma_T, \nonumber \\
Y_i &=& \mathcal{V}_{OM} \left(\frac{x_i^{11}}{L} \right) + i\sigma_i, \hspace{1cm} i=1 \ldots N. \nonumber
\end{eqnarray}
Here $\mathcal{V}$ is the averaged volume of the Calabi-Yau, $L$ is the length of the orbifold direction, $x_i$ is the position modulus of each
individual five-brane satisfying $0 \le x_i \le L$ and $\mathcal{V}_{OM}$ is the average volume of an open membrane
instanton wrapping the same holomorphic two-cycle as the five-brane. The axions $\sigma_s, \sigma_T,$ and $\sigma_i$ arise from various
components of the 3-form of 11-dimensional supergravity and play an important role when considering inflation using a single brane \cite{copeland}, however for the purposes of this note we will
neglect their contribution.
It is convenient for us to write the real moduli in the following way
\begin{eqnarray}
s = S + \bar{S}, \hspace{0.5cm} t &=& T + \bar{T}, \hspace{0.5cm} y_i =Y_i + \bar{Y_i},\\
y&=& \left(\sum_{i=1}^{N} y_i^2 \right)^{1/2} \nonumber
\end{eqnarray}
in terms of which we find the leading order contribution to the K\"ahler potential can be written as follows
\begin{equation}\label{eq:kahler}
K = -\ln\left(s-\frac{y^2}{t} \right)-\ln \left(\frac{dt^3}{6} \right)-\ln \left(i \int_{CY} \Omega \wedge \bar{\Omega} \right),
\end{equation}
where $d$ is the intersection number of the Calabi-Yau, $\Omega$ is the $(3,0)$ form on the Calabi-Yau and we are taking $h^{1,1}=1$ for simplicity.
There are two things of note to mention at this stage. First that this represents the 
leading order expansion of the K\"ahler potential in powers of the strong coupling parameter, which measures the relative size of loop corrections to the four dimensional effective action \cite{decarlos}. Secondly we require that $s \ge y^2/t $ in order for the potential to be real. In fact we will consider large volume compactification where $s >> y^2/t$ as in \cite{becker}.

The superpotential we will use is assumed to arise solely from inter-membrane interactions, where the instantons
wrap the same cycle as the five-branes in the Calabi-Yau. There are also other non-perturbative corrections to the superpotential arising
from the boundary-boundary instantons and also gaugino condensation on the boundary, however we will assume that these are all negligible.
We will also set the charged matter fields to zero. 
Therefore the superpotential arising from the inter-membrane instantons is given by
\begin{equation}
W = h \sum_{i<j} e^{-Y_{ji}}, \hspace{1cm} Y_{ji} = Y_{j} - Y_{i},
\end{equation}
where $h$ is a holomorphic section of a line bundle over the moduli space of complex structure.
This is used to calculate the four dimensional F-term $\mathcal{N}=1$ scalar potential via
\begin{equation}
V =  e^{K} \left(\sum K^{I \bar{J}} D_I W \overline{D_JW} -3|W|^2 \right),
\end{equation}
where we have explicitly set $M_p = 1$. Upon substitution of the expression for the K\"ahler potential and using our large volume approximation, we obtain the 
following form of the scalar potential
\begin{eqnarray}
\frac{V}{e^{K}} &=& G^{\alpha \bar{\beta}} D_{\alpha}W \overline{D_{\beta}W} + Qt \sum_{i,j=1}^N \left(\frac{Qy_iy_j}{Rt} + \frac{\delta_{ij}}{2} \right)\overline{D_iW}D_jW + {}\\
& & {} + \left( \frac{3Q^2}{R}-\frac{2y^2}{Qt} \right)|W|^2, \nonumber
\end{eqnarray}
where we have written $Q=s-y^2/t$ and $R=3Q^2-2y^4/t^2$ for simplicity. The inverse metric $G^{\alpha \bar{\beta}}$ runs over the complex structure moduli, however we
set this contribution to zero using $D_{\alpha}W=0$ as this minimises the energy and also leads to a constraint on the value of $h$ in the superpotential.
Further consideration of energy minimisation leads us to impose the additional condition $D_i W = 0$, which has a direct physical interpretation for the theory.
The K\"ahler derivative acts upon the superpotential as $D_i W = \partial_i W + W \partial_i K$, however the second term is negligible using the supergravity constraint - namely that $s>>1$ or 
large volume compactification -
and the first term represents the variation in distance each brane is from one another. By setting all the branes to be equidistant from one another we ensure that the
first term will also vanish. This also means that the largest contribution to $W$ will come from nearest neighbour instantons and so using $\Delta Y = Y_{i+1}-Y_i$
we find the potentials reduce to
\begin{eqnarray}
V &=& \frac{6}{d(i\int \Omega \wedge \bar{\Omega})} \left(\frac{3Q}{Rt^3}-\frac{2y^2}{Q^2t^4} \right) |W|^2 \\
W &=& h \sum_{i}^{N-1} e^{-\Delta y/2} \nonumber
\end{eqnarray}
As we want to map this picture to that of power law inflation we must ensure that there is no dependence on the inflaton in the pre-factor
term of the scalar potential, which we can achieve by demanding $Qt^2 >> y$ in order for the second term to be negligible. However since we are using the large volume approximation, we see that $Q \sim s$ and $R \sim 3s^2$ therefore the pre-factor term reduces to unity and we are simply taking the scalar potential to
be the modulus squared of the superpotential, multiplied by the K\"ahler factor, $e^K$.
The resultant potential becomes
\begin{equation}
V = \frac{6}{st^3 d (i\int \Omega \wedge \bar{\Omega})} |W|^2,
\end{equation}
which we can simplify further by noting that stabilising the complex structure terms implies that $h = (i\int \Omega \wedge \bar{\Omega})$ and thus
\begin{equation}
V = \frac{6(i\int \Omega \wedge \bar{\Omega})(N-1)^2}{st^3d} e^{-\Delta y/2},
\end{equation}
where we use the real part of the $T$ modulus to define the physical separation of the branes.
Furthermore as we are interested in the dynamics of the fivebranes along the orbifold direction, we only need to include a kinetic term for the $Y_i$ fields in
our four dimensional effective action. As argued in \cite{becker} we can safely assume that the orbifold and the Calabi-Yau 
are of fixed size during inflation as there are no steep potentials for these moduli which couple to the inter-brane scalar fields.
The resultant kinetic term in the action reduces to
\begin{equation}
S = -M_p^2 \int d^4 x \sqrt{-g} K_{i \bar{j}}\partial_{\mu} y^i \partial^{\mu} \bar{y}^j,
\end{equation}
where $i, j$ run from $1 \ldots N$. As we are interested in realising the Assisted Inflation scenario from M-theory it is useful to
rescale the fields to make them canonically normalised, however as there are $N$ equidistant branes it will be easier to consider the collective
centre of mass field $\phi_{cm}$ rather than the individual $\phi_i$. There is a simple relationship between the centre of mass field and the distance
between each brane which allows us to explicitly perform the summation in the superpotential \footnote{The reader is referred to \cite{becker} for a more detailed 
description of the statements here.}.
The end result is that we can express our canonically normalised field as follows
\begin{equation}
\phi = M_p \sqrt{\frac{2N(N^2-1)}{3Qt}} \Delta y,
\end{equation}
where $\Delta y$ represents the fixed distance between each of the fivebranes, and the factor $(N-1)$ arises from explicitly performing the summation
in the superpotential. Inflation in this case will only end once other terms in the superpotential become important, such as the brane-boundary or the
boundary-boundary terms. Once this happens the size of the orbifold will start to grow \cite{becker, buchbinder2}, in turn forcing the gauge coupling on the hidden 
boundary to increase. Eventually we move into the strongly coupled regime which triggers gaugino condensation on the boundary, which in turn can stabilise the orbifold length and the overall volume of the Calabi-Yau \cite{moduli_stabilisation}.

This potential gives rise to power law inflation provided that the number of branes is around $N\sim 89$. Furthermore we see that the scalar index 
$n_s$ is a function of the number of branes, but for the case above turns out to be $n_s = 0.98$ which is slightly blue shifted but yields an almost scale invariant spectrum
in perfect agreement with observational data \cite{data}. 
\section{Inflationary Attractor.}
The construction of the previous section yields the following potential for our canonically normalised scalar field
\begin{equation}
V = \tilde{V_0}(N-1)^2 e^{-A \phi} = V_0 e^{-A\phi},
\end{equation}
where we have introduced the following simplifying notation
\begin{equation}
\tilde{V_0} = \frac{6M_p^4 (i \int \Omega \wedge \bar{\Omega})}{st^3 d}, \hspace{1cm} A=\frac{1}{M_p}\sqrt{\frac{3Qt}{2N(N^2-1)}}.
\end{equation}
Note that we have retained the explicit Planck scale dependence to remind the reader that the field is canonically normalised, however we
are still employing the convention $M_p=1$.
The equation of motion for the inflaton can be derived from the full four dimensional effective action which we minimally couple to an Einstein-Hilbert term, a
simple variation yields \cite{becker}
\begin{equation}
\ddot{\phi} + 3H\dot{\phi} + V' =0,
\end{equation}
whilst the corresponding Friedmann equation is simply
\begin{equation}
H^2 = \frac{1}{M_p^2} \left(\frac{\dot{\phi}^2}{2} + V \right),
\end{equation}
where $H$ is the usual Hubble parameter, primes denote derivatives with respect to $\phi$, dot denotes derivatives with respect to time, and we are assuming a flat FRW (Friedmann-Robertson-Walker) metric.
This potential is actually a special case of power law inflation, and therefore satisfies the slow roll conditions as well as generating the required number
of e-foldings with a flat, scale invariant spectrum.
However for our purposes, is it useful to look at the stability of the solution using the equations of motion to check if this corresponds
to an attractive point in phase space \cite{phase_space}. Phase space trajectories provide useful information about cosmological systems as they are 
independent of any particular initial condition we impose on the theory. Instead they pick out the late time orbits around sink and source points.
We must first rewrite the equations in terms of dimensionless variables, for simplicity we choose $x=\dot{\phi}/M_p^2$ and $y=V/V_0$.
By calculating the value of $\Omega_{\phi} = \rho/(3M_p^2 H^2)$, we see that both $x$ and $y$ belong to the closed set $[0,1]$ which is to
be expected as we are using conventions where the speed of light is equal to unity.
Furthermore it is convenient to use the number of e-foldings $N_e$, as our new dimensionless time variable, where $N_e = \ln(a)$ is a measure of the expansion
of the universe with respect to the scale factor. Also note that $A M_p$ is dimensionless, but we don't write it as a new phase space variable. 
We denote these derivatives by a prime, and see that the equations reduce to the simple forms
\begin{eqnarray}
x^\prime &=& \frac{AV_0y}{M_p\sqrt{V_0 y + M_p^4 x^2/2}}-3x \\
y^\prime &=& -\frac{Ayx M_p^3}{\sqrt{V_0 y + M_p^4 x^2/2}}. \nonumber
\end{eqnarray}
The only fixed point solution can be seen to occur at $x=0, y=0$, which makes the stability analysis relatively simple. This solution would generically be expected
as it corresponds to constant field value at the zero of the potential. In terms of the fivebrane picture this would be where the branes strike the boundary.
We consider perturbations of
our variables about their values at the critical points, $x = x_c + \epsilon$ and $y=y_c + \delta$, and expand the equations to leading order. In order to
linearise the resulting expressions we must assume that $\epsilon^2 << 1$, and we also rescale $\delta$ so that it
can be written $\delta = \gamma^2$. Our final expressions are simply
\begin{eqnarray}
\epsilon^\prime &\sim& \frac{A \sqrt{V_0} \gamma}{M_p} - 3\epsilon \\
\gamma^\prime &\sim& - \frac{AM_p^3 \epsilon}{2\sqrt{V_0}}
\end{eqnarray}
Writing these as a matrix equation, we can integrate out the $N_e$ dependence to find the solution
\begin{equation}
\Psi^{a} = \Psi_0^{a} e^{\mathbf{M}N_e},
\end{equation}
where we have coupled the perturbations into a two component vector, and $\mathbf{M}$ is therefore a two dimensional matrix which
has no explicit dependence upon the number of e-foldings
\begin{equation}
\mathbf{M} = \left( \begin{array}{cc} -3 & \frac{A \sqrt{V_0}}{M_p} \\  -\frac{AM_p^3}{2\sqrt{V_0}} & 0 \end{array} \right).
\end{equation}
The eigenvalues of this matrix allow us to examine the stability of the fixed point in phase space, since a stable solution
requires both eigenvalues to be negative.
Fortunately it is trivial to calculate the eigenvalues of this matrix, which are given by the following expression
\begin{equation}\label{eq:eigen}
\lambda_{\pm} = \frac{-3}{2} \pm \frac{1}{2}\sqrt{9-2A^2M_p^2}
\end{equation}
which is valid only for $N > 1$ and is independent of $V_0$ proving that the initial conditions do not determine the final trajectory through phase space! For the moment we 
will concentrate on the dependence upon the number of five-branes by taking the large $N$ limit and assuming that the Calabi-Yau data is fixed. This is a reasonable 
assumption because we know that in this limit $A$ is usually small. Upon substitution we find the eigenvalues simplify to
\begin{equation}
\lambda_{\pm} \sim \frac{-3}{2} \pm \frac{1}{2}\sqrt{9-\frac{3st}{N(N^2-1)}} 
\end{equation}
Therefore we find two negative eigenvalues, although one of them can be close to zero which indicates an almost flat direction in phase space
\footnote{Of course a zero mode implies that the determinant is zero which complicates the analysis, however this can only occur when we take the $N \to \infty$
limit of the theory which we regard as being unphysical as we are then demanding an infinite number of branes along a finite interval.}.
This turns out to be a stable attractor solution, and as such implies that all trajectories converge toward this fixed point.

Referring back to (\ref{eq:eigen}) we see that the stability of the solution is dependent upon
the region of moduli space we are probing. As $d$ is typically of order unity, we can
neglect its contribution to the problem and instead concentrate only upon the $s$ and $t$ moduli.
The result is that the second term in the square root is dependent upon the volume and orbifold moduli, and could provide the dominant contribution in the square root 
provided that the Calabi-Yau data is fine tuned - or by considering a smaller number of five-branes along the interval. In any case, this complexifies the
eigenvalues and therefore alters the trajectories through phase space.
However because the real part of the eigenvalues are negative, we find oscillatory behaviour about the fixed point solution. Thus we may envisage a spiralling trajectory
in phase space which ends at our fixed point. Generally we would expect this latter solution to be unphysical because we must demand $A<<1$ for inflation to occur.
The significance of this is that inflation in this scenario has a solitary fixed point, which can be regarded as being stable independent of the
exact choice of parameters.

As is usual in inflationary scenarios we are able to study the evolution of the inflaton field
in various domains. Initially the universe is dominated by the potential energy of the field,
where the inflaton is at an arbitrary point on the steep potential but having an initial
velocity of zero. The equations of motion in this epoch reduce to
\begin{eqnarray}
0&\sim&\ddot{\phi}+V^{\prime} \nonumber \\
H^2 &\sim& \frac{V}{M_p^2}.
\end{eqnarray}
Choosing the initial conditions $\dot{\phi}=0$ and $V(\phi_0)=V(\phi(t=0))$ we 
can obtain the initial solutions for the scalar field and the scale factor
\begin{eqnarray}
a(t)&=& \exp \left(\frac{1}{M_p} \int_0^t dt \sqrt{V} \right) \nonumber \\
\phi(t) &=& \sqrt{2} \int_0^t dt \sqrt{V(\phi_0)-V(\phi)}. 
\end{eqnarray}
These solutions only hold provided that the velocity of the scalar field is extremely small,
and should be interpreted as initial constraints upon the inflaton.
The field starts at a large initial value of the potential and rolls very slowly down the slope, due
to the large friction term coming from the Hubble parameter. The universe will be expanding
slowly at this point allowing for a suitably large number of e-foldings, eventually reaching an epoch where the kinetic
energy totally dominates the evolution. When the field reaches this point we must approximate the equations 
of motion by
\begin{eqnarray}
\ddot{\phi} &\sim& -3H\dot{\phi} \nonumber \\
H^2 &\sim& \frac{\dot{\phi}^2}{2M_p^2},
\end{eqnarray}
which implies that the scale factor can be approximated by
\begin{eqnarray}
a(t)&=& \exp \left(\frac{1}{\sqrt{2}M_p} \int_{t'}^t d\phi \right) \nonumber \\
&=& \exp \left(\frac{\phi(t)-\phi(t')}{\sqrt{2}M_p} \right),
\end{eqnarray}
where $t'$ represents the initial time when the kinetic terms will dominate over the potential
terms. These equations can be regarded as the asymptotic late time approximate solutions for the 
evolution of the universe.
\section{Instanton corrections to the Inflaton potential.}
The analysis thus far has effectively neglected any higher order non-perturbative corrections to the scalar potential, which has
allowed us to map the cosmological solution to that of the Assisted Inflation scenario \cite{becker}. In this section we will look at
some of the possible corrections to the four dimensional action and their ramifications for cosmological evolution which were not considered in the
previous model.
In particular we want to see whether including the boundary-brane correction prevents us from achieving slow roll in this model.
Our starting point, once again, will be the K\"ahler potential given by (\ref{eq:kahler}) which allows us to calculate the
K\"ahler metric (see Appendix) for the four dimensional theory.
The resulting $\mathcal{N}=1$ effective action will be given by
\begin{equation}
S = -M_p^2 \int d^4 x \sqrt{-g} \left(\frac{R}{2}+K_{i\bar{j}}\partial_{\mu} \phi^i \partial^{\mu} \phi^{\bar{j}} + V \right),
\end{equation}
where $V$ is the usual scalar potential and the label $i$ runs over the chiral superfields $(s,t,y)$.
\subsection{Boundary-Brane contributions.}
The approximation made in \cite{becker} was that the superpotential is generated entirely by the inter-brane instantons, and that
each brane is equidistant implying that only the nearest neighbour instantons are relevant. We would like to go beyond this
to calculate the next most important terms arising from the boundary-brane instanton contribution. Becker et al. \cite{becker} argued that inflation would
generically end at the point where the competing terms in the superpotential became equal. This is because the contribution from the boundary instanton
term is sufficiently large to make the slow roll parameters always be $\mathcal{O}(1)$. Therefore we expect any model with this interaction included
from the start would generally not allow for inflation, as it implies that the relative size of the orbifold is small. However, the form of the potential
at the end of inflation is relevant for calculating observables and may also give us information about reheating.

Recall from our set-up that the compactified theory
has a visible and hidden sector located at each of the orbifold fixed points with a $\mathbb{Z}_2$ symmetry relating each of them. Thus
we can effectively ignore one of the boundaries, and use the symmetry of the problem to determine the full dynamical behaviour. 
Therefore we will take the superpotential to be of the form
\begin{equation}
W = h \sum_{i < j} e^{-(Y_j-Y_i)} + h\sum_{i=1}^{N/2} e^{-Y_i}.
\end{equation}
The first term comes from instantons wrapping the same holomorphic two cycle as the branes, whilst the second term is due to
the distance of each brane from one of the orbifold fixed points. Note that the form of the second term implies that the
leading branes, namely those closest to the fixed points, will provide the largest contribution to the superpotential.
The sum is over the $N/2$ branes which are in this section of the orbifold, and for simplicity we will assume that the pre-factor $h$ is the same for both
kinds of contribution to the superpotential.

We now re-calculate the scalar potential using the K\"ahler metric and the superpotential shown above. It is useful to consider the supergravity approximation
where $st >> y^2$, and also that there is an equidistant brane configuration. Thus the five-branes, placed arbitrarily along the orbifold,
would tend to their equilibrium equidistant configuration in the 
absence of any potentials. As usual this energy minimisation condition dramatically simplifies the problem, since we can set $Y_j-Y_i=\Delta Y$
to be the inter-brane distance, and this also allows us to take nearest neighbour interactions as the dominant contribution to the inter-brane instanton term.
With this in mind, we can explicitly calculate the first term in the superpotential to be the following
\begin{equation}
W \sim h \left( (N/2-1)e^{-\Delta Y} + \sum_{i=1}^{N/2} e^{-Y_i}\right).
\end{equation}
After dropping the complex structure terms to ensure a minimal energy configuration, we are left with the following expression for the scalar potential
\begin{equation} \label{eq:full_potential}
V \sim e^K h \bar{h} \left(C^2 + C \sum_i (e^{-Y_i} + e^{-\bar{Y}_i})\left\lbrace1+y\left(2-\frac{1}{t} \right)\right\rbrace + \sum_i \sum_j e^{-(Y_i+\bar{Y}_j)}\left\lbrace 1+\frac{s}{2}+y\left(2-\frac{1}{t}\right)\right\rbrace \right),
\end{equation}
where we have introduced the notation $C=(N/2-1)e^{-\Delta Y}$ for simplicity. Clearly when considering only the leading term in this expression we will
recover the usual potential for the Assisted Inflation solution. 
The final double summation in the potential arises due to cross terms, however when we use the real modulus $y$
we see that the contribution from this term will be sub-leading and so we can neglect its contribution.
However the inclusion of boundary instantons implies certain consistency conditions upon the potential. As the five-branes move in their potential there will
eventually be a time when some of the boundary instantons are of a similar magnitude to the inter-brane distance.
For the purpose of this section we will
assume that we can safely include the boundary instantons associated with the first and second branes - as measured from our chosen fixed point.
This gives us the approximate potential
\begin{equation}\label{eq:approx_potential}
V \sim e^K h\bar{h} \left(C^2+C[e^{-Y_1}+e^{-\bar{Y}_1}+e^{-Y_2}+e^{-\bar{Y}_2}] +\frac{s}{2} [e^{-(Y_1+\bar{Y}_1)}+e^{-(Y_1+\bar{Y}_2)}+e^{-(Y_2+\bar{Y}_1)}+e^{-(Y_2+\bar{Y}_2)}]\right),
\end{equation}
when we impose the additional constraint $1>>2y-2/t$. This relates the size of the orbifold to the scalar field through the relation $t >> 1/y$, where we want $t > 1$ in agreement
with supergravity constraints.
It remains to calculate the kinetic terms for the four dimensional action using the K\"ahler potential, noting that the inflaton field, $y$, is the only dynamical
field in our theory. Once again it is necessary to switch to the canonically normalised field rather than the superfield, and so we employ the following transformation
\begin{equation}
\phi^i = \frac{2y^i}{\sqrt{st}}, \hspace{0.5cm} \Delta \phi^i = \frac{2\Delta y^i}{\sqrt{st}},
\end{equation}
In fact it will be more convenient to consider an alternative field, namely the centre of mass field for all $N$ branes as in the Assisted Inflation scenario.
Denoting this field by $\psi$ we find an expression for $\Delta y$ in terms of this new field which we also canonically normalise
\begin{equation}
\Delta y = \sqrt{\frac{3st}{N((N/2)^2-1)}} \psi \hspace{0.3cm} = A\psi.
\end{equation}
Thus far our argument has run along the same lines as \cite{becker}, however we must also consider the $Y_i$ field which corresponds to the boundary instanton correction.
We note that each of the branes are equally spaced, thus the $i$th brane will be located at the position $y_i = y_1 + (i-1) \Delta y$, where the last term will
contribute a constant piece to the potential. It looks as though we will need an extra kinetic term for the $y_1$ field, however we argue that this can be related to the centre of
mass field in the following way.
Note that  $\psi$ is an increasing function, since as the branes move along the orbifold direction their centre of mass position will change. In fact
when $\psi$ is of order of the orbifold length we know that inflation will naturally terminate due to the argument in \cite{becker}.
Clearly we can make headway here by postulating that at the start
of inflation the fivebranes are located at some point around $x=L/2$ where $L$ is the natural length of the orbifold. The distance spanned by the branes from the
centre of the orbifold is simply $x \sim (N/2) \Delta y = (N/2) A\psi$ which in turn means that the first brane is located at a distance $y_1 \sim \mathcal{V}_{OM} - (N/2)A\psi$
from the orbifold boundary \footnote{Recall that we are only considering a single boundary for simplicity.} when we use the definition of the superfield. We can write this in 
a simpler fashion by noting that $t \sim 2 \mathcal{V}_{OM}$, where $\mathcal{V}_{OM}$ is the averaged volume of an open membrane
instanton wrapped on a two-cycle between the two boundaries. This means
that we can still use a single scalar field to drive inflation in our effective action.
The instanton contribution to the superpotential becomes
\begin{eqnarray}
W_{inst} &\sim& \sum_i^{N/2}  e^{-1/2[y_1 +(i-1)A\psi]} \\
&\sim & e^{-1/4[t-A\psi(2+N)]} \sum_i^{N/2} e^{-Ai\psi/2} \nonumber \\
&\sim& e^{-1/4[t-A\psi(N+2)]} \left(\frac{e^{-A\psi/2(N/2+1)}-e^{-A\psi/2}}{e^{-A\psi/2}-1} \right) \nonumber,
\end{eqnarray}
and so we write the full potential (\ref{eq:full_potential}), to leading order, as follows
\begin{equation}
V \sim V_0 e^{-A\psi}\left((N/2-1)^2+2(N/2-1) e^{A\psi/2} e^{-1/4(t-A\psi(N+2))}\left \lbrace \frac{e^{-A\psi(N/2+1)/2}-e^{-A\psi/2}}{e^{-A\psi/2}-1} \right \rbrace \right),
\end{equation}
where we must also assume that $V_0$ is approximately constant during inflation as there are no steep potentials for the $S$ or $T$ moduli. Recall that the expression for $V_0$ is given by
\begin{equation}
V_0 = \frac{6(i \int \Omega \wedge \bar{\Omega})}{sdt^3}.
\end{equation}
Now taking our approximations into account by dropping all instanton terms which are comparable with the inter-brane distance, i.e only keeping $y_1$ and $y_2$ boundary instantons,
leaves us with the following
\begin{equation}\label{eq:inflaton_potential}
V(\psi) \sim V_0 e^{-A\psi} \left((N/2-1)^2 + (N-2)e^{1/2A\psi(1+N)}e^{-t/2} \left\lbrace 1+e^{-A\psi}\right\rbrace \right).
\end{equation}
Thus our inflaton potential is explicitly dependent upon the number of branes along the section of the orbifold, and the length of the orbifold itself through the volume of the open membrane
instanton. Taking
$L$ to be large automatically implies that the correction terms are negligible and so the potential reduces to that of the previous section. If we search for turning points
we find the only minimum at $\psi \sim t/(AN)$ in the large $N$ limit. This corresponds to the physical distance $y_1 = 0$ implying that the leading brane has coalesced with the boundary, possibly
via a small instanton transition in which case we will have to also take into account the modulus associated with the gauge bundle.
In any case, physically the fivebranes will always move toward the boundary brane in this potential, thus we do not expect there to be any non-trivial dynamics
as in \cite{copeland, decarlos} where a solitary brane exhibited a bounce solution along the orbifold.

We now want to see if the boundary-brane instanton corrections significantly alter the dynamics of the inflaton field. It is a general fact that any potential arising from
non-perturbative effects will be 
exponential, implying that inflation will be difficult to achieve because the inflaton rolls too rapidly toward the minimum
Assisted inflation is able to side-step this problem because the large number of fields/ branes enhance the contributions to the friction term
in the inflaton equation of motion. In this instance the extra corrections from the first and second instanton contributions
will make the mapping to power law inflation difficult. As such we proceed with a more typical inflationary analysis.
Inflaton potentials are generally characterised by various slow roll parameters, which must be substantially smaller than unity in order for inflation to
occur. We use the following definitions of the parameters using the reduced Planck mass as is customary in string theory
\begin{equation}
\epsilon = \frac{M_p^2}{2} \left(\frac{V^\prime}{V} \right)^2 \hspace{0.5cm} \eta = \frac{M_p^2 V^{\prime \prime}}{V}, \hspace{0.5cm} \zeta = M_p^4 \frac{V^{\prime \prime}V^\prime}{V^2},
\end{equation}
where primes denote derivative with respect to the inflaton field. 
Many of the problems related to embedding inflation into string theory arise because $\eta$ is generically too large.
In order to proceed we first choose to take the large $N$ limit of (\ref{eq:inflaton_potential}). This is not an unreasonable approximation
as we know that increasing the number of branes increases the Hubble friction term. We will also assume that $\mathcal{V}_{OM}$ is fixed during inflation allowing the boundary-instanton terms to be included \footnote{Thanks to Axel Krause for pointing out an algebraic error in a previous draft of this note.}.
Of the slow roll parameters it is the first two which are the most important to satisfy, upon calculation we find that the solutions are given by
\begin{eqnarray}
\epsilon &=& \frac{A^2e^{2A\psi}}{2}\left(\frac{e^{-A\psi}(2-N)+e^{A\psi/2[N-1]-t/2}(2N-2)+e^{A\psi/2[N-3]-t/2}(2N-6)}{N-2+4e^{A\psi/2[N-1]-t/2}+4e^{A\psi/2[N+1]-t/2}}\right)^2 \nonumber\\
\nonumber \\
\eta &=& A^2 \left(\frac{N-2+e^{A\psi/2[N+1]-t/2}(N-1)^2+e^{A\psi/2[N-1]-t/2}(N-3)^2}{N-2+4e^{A\psi/2[N-1]-t/2}+4e^{A\psi/2[N+1]-t/2}}\right).
\end{eqnarray}
We see that there is now non-trivial dependence upon the length of the orbifold in the inflaton potential which complicates our analysis.
Because the potential is exponentially decreasing, we need to ensure that the parameters at the top of the potential are amenable to the slow roll analysis - otherwise inflation will not
be possible. Typically $\eta$ is the hardest condition to satisfy so we choose to look at that expression. In the large $N$ limit, and in Planck units, we can Taylor expand the parameter around the origin to find
\begin{equation}
\eta(0) \sim A^2 N \left(\frac{1+2Ne^{-t/2}}{N+8e^{-t/2}} \right),
\end{equation}
which we know must satisfy $|\eta| << 1$. In fact this imposes a constraint upon the number of five-branes in the theory as 
can be seen using the full form of the Taylor expansion and assuming typical values for the Calabi-Yau \cite{becker, desitter}. We find
that $N \ge 55$ to satisfy the initial conditions for inflation as can be seen in Figure 1.
\begin{figure}[ht]
\begin{center}
\epsfig{file=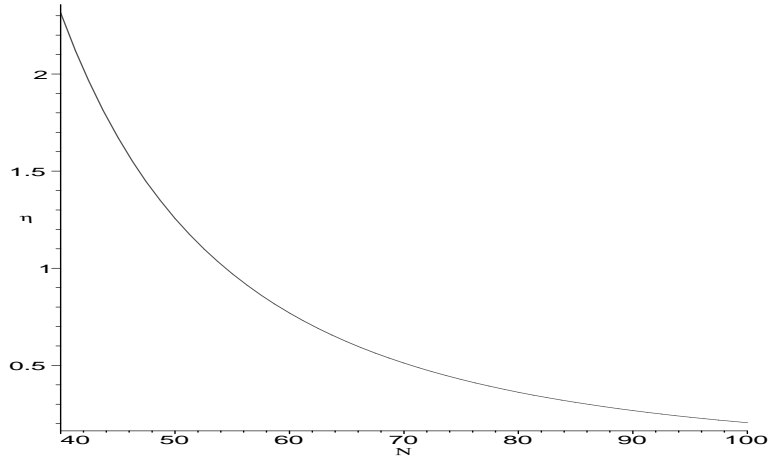, width=10cm,height=6cm}
\caption{Initial value of $\eta$ as a function of $N$, note that we must ensure $\eta <<1$ in order for inflation to occur. This puts a bound on
the smallest number of branes that we can include in the theory. We have assumed that $t=14$ in order to generate this plot.}
\end{center}
\end{figure}
Rather than solving this for $N$, as we know that increasing the number of branes will increase the friction and slow the inflaton, we instead wish to turn this into a constraint upon the length of the orbifold. After some algebra we find that in order for the slow roll parameter to be satisfied
we need
\begin{equation}\label{eq:Lbound}
N-2+4(X_{+}+X_{-}) >> A^2(N-2+(N-1)^2X_{+}+(N-3)^2X_{-}),
\end{equation}
where we have introduced the notation $X_{+}=e^{\frac{A\psi(N+1)-t}{2}}$, $X_{-}=e^{\frac{A\psi(N-1)-t}{2}}$. Note that
this is valid for any $\psi$, however as the inflaton rolls down its potential we require larger and larger values of $t$ to satisfy the slow roll condition. But since $t$ is not
assumed to be dynamical we see that this is an unphysical constraint.
As we know that $N \ge 55$ for the slow roll to be satisfied, we can confidently use the large $N$ approximation in our
constraint equation, which simplifies to
\begin{equation}
N^2(N+8e^{(AN\psi-t)/2}) >> 12st(1+2Ne^{(AN\psi-t)/2}),
\end{equation}
implying that inflation ends when
\begin{equation}
\psi_e = \sqrt{\frac{N}{12st}}\left(t+2\ln\left(\frac{N^3-12st}{8N(N-3st)}\right) \right).
\end{equation}
Again taking typical values for our Calabi-Yau we see that this equation breaks down and becomes negative above $N \sim 51$ implying that inflation will not be possible in 
this instance. In the region where inflation could occur, plots of the slow roll parameter show that $\eta$ is always larger than unity again implying that inflation
is not possible.

By studying the solution space of $\eta$ for finite $t$ we see that it is always large, regardless of how much fine tuning we impose. This implies
that we will not have an inflationary phase as the inflaton rolls too rapidly. We should check how many e-foldings can be generated with this instantonic
contribution, because we may have some form of Assisted Hybrid Inflation. The number of e-folds is given by the usual integral
\begin{equation}
N_e = -\int _{\psi}^{\psi_e} \frac{V}{V^{\prime}} d\psi,
\end{equation}
where the upper integration limit corresponds to the point at which the field crosses the horizon. We can numerically integrate the equation by selecting various values for the parameters, however the general result is bleak. Taking $t$ to be extremely large can lead to the required number of e-folds as we can neglect all the instanton contributions. However this just reproduces the minimal Assisted Inflation case where the correction term vanishes. In order to see it's effect on the solution we need to
take $t$ to be small. However the result is that the field only rolls for a short distance and contributes virtually nothing to the number of e-folds. This is to be expected as \cite{becker} used this 
as an argument for the mechanism to end inflation and we have shown this to be correct. This is suggestive that the orbifold must be sufficiently large, and hence
gives credence to the large extra dimensions scenario.

\subsection{Inter-Brane Coupling.}
We also consider what happens when we relax the assumption made about the nearest neighbour interactions. In general we would expect the fivebranes to become equidistant
as this minimises the total energy of the system, however at the start of inflation the inter-brane instantons will all be significantly smaller than the boundary-brane 
instantons and thus dominate the superpotential. In \cite{becker} the assumption was that only nearest neighbour brane interactions should be included, however if we are to ensure the energy scale is below the Planck scale at the start of inflation, we should investigate the strongly coupled brane system. We again assume that the branes have already moved to their equilibrium distance $\Delta Y$, but this time we wish to sum over all the interbrane interactions. This can be done by re-writing the superpotential in the following manner
\begin{equation}
W = h \sum_{k=1}^{N-1} \sum_{i=1}^{N-k} e^{-k \Delta Y},
\end{equation}
where $k$ corresponds to the order of the interaction. For example $k=1$ is just the nearest-neighbour solution we have already discussed, $k=2$ includes the contribution
from the next to nearest neighbour interactions etc etc. Thus we can include all the possible instanton contributions between the branes in a simple manner.
These higher order terms will quickly become sub-dominant and can be dropped provided we also drop the boundary-brane
interaction. In terms of our canonically normalised centre of mass field $\psi$ we write the potential generated from all these interactions by explicitly carrying out
the double summation - bearing in mind that $A$ is now defined as in the original Assisted Inflation scenario
\begin{equation}
V = \frac{V_0 e^{-A\psi} (N-1)^2}{(e^{-AN\psi/2}-1)^4}(1-e^{-AN\psi/2})^2
\end{equation}
The terms arising from the inclusion of other inter-brane interactions are exponentially suppressed as we would expect, however note that there is now a singularity in the 
potential at the origin - namely when $\psi=0$. This occurs due to the strongly coupled interactions between the fivebranes, which will rapidly repel one another.
We will begin by considering some special cases of the potential, beginning with the $k=1$ contribution only which is the solution presented in \cite{becker}. In terms of our slow roll parameters we find
the following solutions
\begin{equation}
\epsilon = \frac{A^2}{2}, \hspace{0.5cm} \eta= A^2, \hspace{0.5cm} \zeta= -V_0 A^3 e^{-A\psi}(N-1)^2,
\end{equation}
and so to obtain inflation we must demand that $A^2 << 1$ or in terms of the dependent variables, that $N(N^2-1) >> 3st$ which is just what we expect from the power law solution.
Now let us consider the case when $k=2$ so that we are including the contributions from the next to nearest neighbour interactions. The potential can be written as follows
\begin{equation}
V = V_0 e^{-A\psi} \left(N-1+Ne^{-A\psi/2}-2e^{-A\psi/2} \right)^2,
\end{equation}
and the corresponding slow roll parameters can be calculated to yield
\begin{eqnarray}
\epsilon &=& \frac{A^2}{8}\left(\frac{2N-2+3Ne^{-A\psi/2}-6e^{-A\psi/2}}{N-1+Ne^{-A\psi/2}-2e^{-A\psi/2}} \right)^2 \\
\eta &=& \frac{A^2}{2}\left(\frac{4N-4+9Ne^{-A\psi/2}-18e^{-A\psi2}}{N-1+Ne^{-A\psi/2}-2e^{-A\psi/2}} \right) \nonumber \\
\zeta &=& \frac{-V_0A^3e^{-A\psi}(2N-2+3Ne^{-A\psi/2})(4N-4+9Ne^{-A\psi/2}-18e^{-A\psi/2})}{N-1+Ne^{-A\psi/2}-2e^{-A\psi/2}}. \nonumber
\end{eqnarray}
As usual $\eta$ is the problematic parameter in supersymmetric theories, and so we will use this to determine the initial conditions for inflation. Note that now all the
slow roll terms are $\psi$ dependent, breaking the power law behaviour.
We demand that $\eta << 1$ at the initial point of inflation, which translates into the constraint
\begin{equation}
\frac{N(N^2-1)(2N-3)}{(2N-7)} >> 3st,
\end{equation}
which is similar to the condition arising from the $k=1$ case. Note that on the left hand side we now have a parameter greater than unity
implying that we can reduce the number of five-branes and still satisfy the slow roll condition. In fact numerics show that we can get away with $N \sim 50$ and still satisfy the 
slow roll conditions. These equations also tell us when inflation ends, and after rearranging the equation for $\eta$ we find
\begin{equation}
\psi_e = \frac{2}{A}\ln \left(\frac{(N-2)(2-9A^2)}{2(N-1)(2A^2-1)} \right)
\end{equation}
However since $A^2 << 1$ this suggests that $\psi_e$ will have no real solution, which is clearly unphysical and therefore we can never obtain the required number of e-foldings - which can be 
verified using numerical integration provided we specify the Calabi-Yau data.
In fact we see that the number of efoldings increases with $N$ up to around $N=40$ before becoming imaginary.
However the maximum number is only around $17$ which is far short of the required amount. This was determined using the typical Calabi-Yau data in \cite{becker} in order to
provide a natural extension of their Assisted Inflation scenario.

In the general case when we include all possible inter-brane instanton contributions we find an additional problem, namely small field inflation is not satisfied because the potential diverges near
$\psi = 0$ and therefore the slow roll conditions will be violated. As the field evolves it eventually reaches a stabilisation point where the potential is sufficiently smooth and the slow
roll parameters are all significantly less than unity, however the field at the end of inflation is given by
\begin{equation}
\psi_e (\pm) = \frac{2}{AN} \ln \left(\frac{2(A^2N(N-1)+A^2-2)}{2A^2(1-N)-4 \pm 2\sqrt{2}AN} \right),
\end{equation}
which is extremely small regardless of the value of $N$ we choose. Thus we again find that inflation is not possible. The physical solution here is with
the positive sign. It could be argued that the new potential wants to pick out
specific Calabi-Yau, however we can again investigate this numerically for a range of $s, t$ and $N$ - which are the parameters defining the theory. The general conclusion however remains unchanged. For
example in table 1 we show the number of e-foldings ($N_e$) as a function of $N$ for a fixed value of the orbifold size, $t=14$ - which we 
regard as being a typical solution, where we take the initial value of the field to be zero.
\begin{center}
\begin{tabular}{c|c}
$N$ & $N_e$ \\
\hline
80 & 0.0597 \\
100 & 0.0614\\
200 & 0.0683 \\
500 & 0.0766 \\
1000 & 0.0807 \\
10000 & 0.0853
\end{tabular}
\end{center}
\begin{center}
Table 1.
\end{center}
The results for other values of $t$ are similar, as can easily be checked using the definition of the superfields in the
first section and \cite{desitter}. The general trend implies that for larger $t$ we get a significantly smaller number
of e-foldings and therefore require an extremely large number of branes. It turns out to be more favourable to consider 
a small Calabi-Yau, however we still cannot obtain the required 60 e-folds even with a large number of branes. We are forced
to conclude that Assisted Inflation is not possible with these corrections.
\subsection{Higher Order corrections.}
One may also consider other correction terms to the four dimensional effective action. For example we may also include the terms in the superpotential coming from the boundary-boundary
instantons which we may expect to become important when the five-branes are spread over the orbifold, or if we take the orbifold size to be small. However, this term was argued to be negligible
as it cancelled out with the gaugino condensation corrections on the boundary \cite{becker}. We could also include corrections to the K\"ahler potential, such as higher order curvature corrections
due to $R^4$ terms in the action. These corrections were computed explicitly in \cite{anguelova} and were found to be linear in the inverse volume of the Calabi-Yau manifold
\begin{equation}
\delta K \sim \frac{b_2 \kappa^{4/3}\chi}{6s} = \frac{B}{s}
\end{equation}
where $b_2$ is defined in Appendix A, $\chi$ is the Euler number of the Calabi-Yau and $\kappa$ is the gravitational coupling constant.
We proceed as before to calculate the new K\"ahler metric and inverse which will allow us to calculate the
scalar potential,
\begin{equation}
V \sim \frac{6W\bar{W}Ae^{-B/s}}{sdt^3(i \int \Omega \wedge \bar{\Omega})} \left(t^2(3t-2B)+\frac{2y^4(B-t)-6y}{s^2}+2 \right)
\end{equation}
However, this procedure is clearly complicated by the fact that there is now an explicit exponential potential for $s$ which implies that the Calabi-Yau volume can no longer be regarded as being
static as the branes move along the orbifold direction. In order for us to retrieve a simple analytic potential for the inflaton we need to ensure that the volume modulus is stabilised,
so that the Calabi-Yau doesn't decompacitfy. This may be achieved \cite{anguelova} by including superpotential terms arising from perturbative effects and/or gaugino condensation, however 
these contributions will complicate the Assisted Inflation solution and it remains an open problem to embed this into a more realistic compactification. They did, however, show that
once the moduli are stabilised we are left with a weakly de-Sitter solution with a vanishingly small cosmological constant. It may be more pertinent to study single brane inflation \cite{buchbinder, copeland, lukas}
in this model rather than the multi brane inflation we have looked at in this note.
\section{Discussion.}
In this note we have looked at the leading order correction terms to the superpotential arising
from the reduction of Heterotic M-theory down to five dimensions. The extra term arising from
boundary-brane instantons appears to prevent inflation from occurring when we use the centre of mass
of the $N$ branes as the inflaton. In order for us to include the boundary interactions we must
assume that the branes are initially distributed in such a way that some of them are located
near to the boundary. An alternative description is that we are allowing the orbifold length
to be smaller than in \cite{becker}. In any case, the new configuration does not give rise to
inflation. Thus we see that minimally coupled multi field inflation in Heterotic M-theory requires a large extra
dimension.
From an energy point of view the inclusion of this correction term
does not correspond to the minimum energy configuration as discussed by Becker et al \cite{becker} and so
one may argue that it is always a negligible contribution once we take the orbifold size 
to be large in agreement with the supergravity constraint. However from a phenomenological 
perspective this is restrictive on the inflationary models which we can consider, and therefore
suggests that inflation is not something that can be generically constructed in M-theory.
One may also argue that the reason inflation is not realised in this simplistic
scenario is because we are trying to use the centre of mass field as our inflaton. It would be
useful to analyse this set-up without making this assumption as in the general assisted inflation
scenario \cite{assisted_inflation}, although we would generally expect this solution to be complicated
as there will be additional summation over each of the individual $y_i$ fields. 

The additional contributions from nearest neighbour instantons play a more important role than first realised, as their
combined interaction forces the inflaton potential to be infinite at the origin. This complicates the slow roll analysis as the field will
initially be rolling very rapidly and therefore we need to ensure the friction term is sufficiently large to prevent rapid inflation. This is true
for general $N$. If we relax this assumption to include contributions from most, but not all of the instantons, then the 
slow roll conditions can be satisfied by taking a smaller value of $N$. This is saying that the extra interaction between the branes
contributes to the friction term in the equation of motion, and therefore acts to flatten the effective potential. Of course, these extra
contributions will fall off rapidly as the branes move along the orbifold and therefore the friction term will be reduced allowing the inflaton to roll
faster and this is what prevents us from obtaining the required number of e-foldings. Thus the general conclusion we reach is that multi-brane inflation
in Heterotic M-theory is not viable, with the sole exception of the standard Assisted scenario \cite{becker}.

One may have expected that there would be some form of moduli inflation \cite{moduli_inflation} scenario here, where a shrinking orbifold size
could drive inflation in a manner similar to that occurring in type II string theory. Leaving aside any comments about the validity of this approach,
we in any case find that the canonical potential consists of double exponentials making it extremely flat but failing to solve the slow roll constraints for
finite size. The problem here is that we cannot neglect the boundary-boundary instantons in this approach which will have a repulsive effect, and eventually force the
orbifold length to grow larger until it stabilises. It would be interesting to study whether higher order corrections to the K\"ahler potential could satisfy the conditions
for inflation in this instance. Furthermore it would be interesting to study the production of cosmic strings in this scenario, in the same vein as the recent work in \cite{buchbinder2}. In any instance, embedding inflation and subsequent reheating into this theory is still an open subject worthy of investigation.
\begin{center}
\textbf{Acknowledgements.}
\end{center}
This work was in part supported by the EC Marie Curie Training Network MRTN-CT-2004-512194.
JW is supported by a QMUL studentship and wishes to thank the following people for their
invaluable assistance and comments during the writing of this note. S. Thomas, P. Moniz,
C. Papageorgakis, J. Bedford, D. Mulryne,  M. Fairbairn and N. Pidokrajt.
\begin{appendix}
\section{K\"ahler metric}
The K\"ahler potential including correction terms is given by the following expression \cite{becker, anguelova}
\begin{equation}
K=-\ln\left(s-\frac{y^2}{t}\right)-\ln\left(\frac{dt^3}{6}\right)-\ln\left(i\int \Omega \wedge \bar{\Omega}\right)-\frac{A}{s},
\end{equation}
from which we can calculate the associated K\"ahler metric through $K_{a\bar{b}}=\partial_a \partial_{\bar{b}}K$. The metric can be written as follows
\begin{equation}
K_{I\bar{J}} = \frac{1}{Q^2}\left( \begin{array}{ccc} 1-\frac{2AQ^2}{s^3} & \frac{y^2}{t^2} & \frac{-2y}{t} \\ \frac{y^2}{t^2} & \frac{2y^2Q}{t^3}+\frac{y^4}{t^4}+\frac{3Q^2}{t^2} & -\frac{2Qy}{t^2}-\frac{2y^3}{t^3} \\ 
-\frac{2y}{t} & -\frac{2yQ}{t^2}-\frac{2y^3}{t^3} & \frac{2Q}{t}+\frac{4y^2}{t^2}  \end{array}\right),
\end{equation}
where we are using the following abbreviations
\begin{equation}
Q= s-\frac{y^2}{t}, \hspace{1cm} A = \frac{b_2\kappa^{4/3}\chi}{6},
\end{equation}
with the factor $b_2$ defined as 
\begin{equation}
b_2 =12. 2^6 (2\pi)^3 b_1 T_2 (2 \kappa_{11}^{2/3}).
\end{equation}
Note that $\chi$ is the Euler number of the Calabi-Yau, and we also have
\begin{eqnarray}
T_2 &=&(2\pi)^{2/3}(2\kappa_{11}^2)^{-1/3} \\
b_1 &=& ((2\pi)^4 2^{13} 9)^{-1}. \nonumber
\end{eqnarray}
This arises from higher order corrections in $R$ from the bulk supergravity action as computed in \cite{anguelova}.
It is a straightforward computation to calculate the inverse metric when we ignore the correction term coming from $R^4$ in the supergravity action.
\begin{equation}
K^{I \bar{J}} = \left(\begin{array}{ccc}\frac{3s^2t^2-2y^4}{3t^2}& \frac{y^2}{3}& \frac{y}{3t}(3st-2y^4)\\ \frac{y^2}{3}& \frac{t^2}{3}& \frac{yt}{3}\\ \frac{y}{3t}(3st-2y^4)& \frac{yt}{3}& \frac{3s-y^2}{6}
\end{array} \right)
\end{equation}

\end{appendix}


\begin{thebibliography}{99}
\bibitem{comment}
Note that in the KKLT approach they uplifted to de-Sitter vacua by adding $\bar{D}3$-branes to the four dimensional effective action. This is problematic
because they should be included in the full ten dimensional theory where they break supersymmetry and so we lose control over the compactification.
\end{thebibliography}
\end{document}